\documentclass{iopart}
\usepackage{epsfig}
\usepackage{iopams}
\begin{document}

\title[Black hole MACHO binaries detection]
{Can black hole MACHO binaries be detected by the Brazilian
spherical antenna?}

\author[de Araujo et al]{J C N de Araujo$^{1}$, O D Miranda$^{2}$,
C S Castro$^{1}$, B W Paleo$^{2}$  and~O~D~Aguiar$^{1}$}

\address{$^{1}$Instituto Nacional de Pesquisas Espaciais - Divis\~ao de
Astrof\'\i sica \\ Av. dos Astronautas 1758, S\~ao Jos\'e dos
Campos, 12227-010 SP, Brazil}

\address{$^{2}$Instituto Tecnol\'{o}gico de Aeron\'{a}utica -
Departamento de F\'\i sica \\ Pra\c{c}a Marechal Eduardo Gomes 50,
S\~ao Jos\'e dos Campos, 12228-900 SP, Brazil}

\ead{\mailto{jcarlos@das.inpe.br}}

\begin{abstract}
Different studies show that dark matter of non-baryonic origin
might exist. There have been experimental evidences that at least
one form of dark matter has been detected through microlensing
effects. This form of dark matter is named MACHOs (Massive
Astrophysical Compact Halo Objects). The MACHO collaboration
estimated that the masses of these objects are to be in the range
0.15-0.95 $\rm M_{\odot}$, where the most probable mass is of
0.5~$\rm M_{\odot}$. Some authors argue that MACHOs could be black
holes, and that they could form binary systems, BHMACHO binaries.
As is well known binary systems are sources of gravitational
waves. The Brazilian spherical antenna will operate in the
frequency band of 3.0-3.4 kHz, sensitive to binaries of a pair of
0.5 $\rm M_{\odot}$ black holes just before coalescing. In the
present work we study the detectability of these putative BHMACHO
binaries by the Brazilian spherical antenna Mario Schenberg.
\end{abstract}










\section{Introduction}
The issue related to the dark matter in astronomy has a long
history. In the 1930s, for example, Oort claimed that the total
amount of matter in the Galaxy is greater than the visible matter.
In this very epoch, Zwicky found that the velocity dispersion of
galaxies in a cluster exceeds that expected  for a gravitationally
bound stationary system if the only contribution comes from the
galaxies themselves.

One could think that the invisible matter is made of ordinary
matter, i.e., baryonic dark matter. There are, however,
independent evidences that lead to the conclusion that the dark
matter cannot be only made of baryons.

The big bang nucleosynthesis studies, for example, indicate that
the amount of baryons in the universe is
$\Omega_{B}h^{2}_{100}=0.020\pm 0.002$ (95\% confidence level;
see, e.g., Burles \etal 2001, where $\Omega_{B}$ is the baryonic
density parameter and $h_{100}$ is the Hubble constant in units of
$100\;{\rm km\;s^{-1}\;Mpc^{-1}}$). This last figure takes into
account, obviously, the luminous and the dark baryons.

On the other hand, several different studies of the cosmic
background radiation, in particular those related to the primary
anisotropies, indicate that non baryonic dark matter exists. The
WMAP results show, for example, that $\simeq 80 \%$ of the matter
must be in the form of non-baryonic dark matter (Spergel \etal
2003).

It is, therefore, hard to avoid the conclusion that at least part
of the dark matter present in galaxies and in cluster of galaxies
is in the non-baryonic form.

A question that still deserves to be properly answered has to do
with the very nature of the dark matter, be it baryonic or
non-baryonic.

Generally speaking, it is argued in the literature (see, e.g.,
Peacock (1999) for a review, among others), to be conservative,
that the dark matter could be in the form of the so called brown
dwarfs, if baryonic, and WIMPs (weakly interacting massive
particle), if non-baryonic.

Paczy\'{n}ski (1986a) suggested that with the use of the leasing
effect it would be possible to observe nonluminous matter in the
form of brown dwarfs or Jupiter-like objects. He coined the term
``microlensing" to describe the gravitational lensing effect that
can be detected by measuring the intensity variation of a
macro-image of any number of unresolved micro-images.

Also, the search of light variability among millions of stars in
the Large Magellanic Cloud (LMC) could be used to detect dark
matter in Galactic halo (Paczy\'{n}ski 1986b).

It is worth mentioning that Griest (1991) coined the acronym MACHO
(Massive Astrophysical Compact Halo Objects) to denote the objects
responsible for gravitational microlensing. The name MACHO became
very popular and is widely used to refer to any object responsible
for the microlensing effect, whether these objects are located in
the halo of the Galaxy or not, and regardless of their masses.

Paczy\'{n}ski's idea concerning the microlensing triggered many
groups to search for MACHOs. Just to mention a couple of groups
still active in this new field of astrophysics we have, for
example: OGLE (Udalski \etal 1992), EROS (Aubourg \etal 1993),
MACHO (Alcock \etal 1993), among others. These various groups have
been monitoring tens of millions of stars in the LMC, searching
for light variability.

These various studies show that there is not enough MACHOs in the
Galactic halo to account for its mass. Therefore, there should
exist another kind of dark matter in the halo, most probably in
the form of WIMPs.

Alcock \etal (2000, MACHO Project), for example, based on a
maximum-likelihood analysis obtained that a MACHO halo fraction of
20\% for a tipical halo model with a 95\% confidence interval of
8\%-50\%. Moreover, they obtained that the most likely MACHO mass
is between 0.15 and 0.9 $\rm M_{\odot}$, with the most probable
value being 0.5~$\rm M_{\odot}$.

This MACHO mass is substantially higher than fusion threshold of
0.08 $\rm M_{\odot}$, and therefore should shine in some
electromagnetic frequency band, but there is no evidence for that.
We do not enter into a detailed discussion on this issue here, we
refer the reader to the papers by Nakamura \etal (1997), among
others. We only remark that the MACHOs are least likely to be
either white, red or brown dwarfs.

Nakamura \etal (1997) argued that the MACHOs could well be
primordial black holes. Obviously, it is not possible to form a
black hole of 0.5~$\rm M_{\odot}$ as a product of the stellar
evolution, it must have formed necessarily in the very early
universe (see,e.g., Yokoyama 1997).

Here we are not concerned with the formation mechanism of this
putative 0.5~$\rm M_{\odot}$ black holes. In the present paper we
consider that they exist, and as a result can form binary systems
(the BHMACHO binaries), that, therefore, may generate
gravitational waves (GWs).

The paper is organized as follows. In section 2, we consider the
GWs from the spiralling BHMACHO binaries, in section 3 we consider
the detectability of the BHMACHO binaries by the Brazilian
spherical detector and finally in section 4 we present our
conclusions.

\section{Gravitational waves from spiraling BHMACHO binaries}
\label{sec-2}

\begin{figure}
\begin{center}
\epsfig{figure=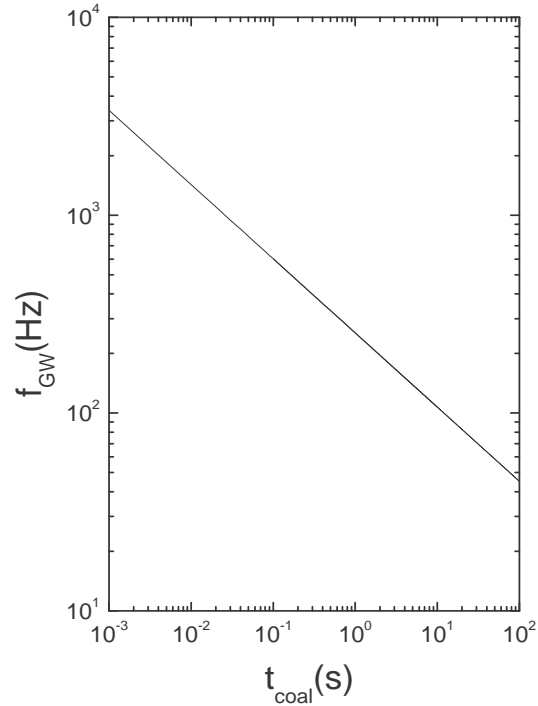,angle=360,height=11cm,width=8cm}
\caption{\label{fig1} The frequency of the GW, $f_{GW}$, as a
function of the coalescing time, $t_{coal}$, for a pair of
0.5~$\rm M_{\odot}$ BHMACHOs.}
\end{center}
\end{figure}

Binary star systems are well-known sources of GWs that should be
detected either by earth or space based GW detectors. Here we are
interested in studying if the Brazilian spherical detector ``Mario
Schenberg" is able to detect BHMACHO binaries.

Due to the fact that the Brazilian antenna will operate in the
frequency band of 3.0-3.4 kHz, GWs emitted, during the periodic or
spiraling phase of the evolution of binary systems, can only be
detected if the components of the binaries are compact objects of
sub-solar masses.

Since the MACHOs could be black holes of 0.5~$\rm M_{\odot}$, the
BHMACHO binaries could in principle be detected by the Brazilian
antenna.

Obviously the other GW detectors (the interferometers and the
bars) will also see such putative BHMACHO binaries. As the
interferometers are sensitive to GWs in the frequency band of 10
Hz-10 kHz, and the bars to $\sim$ 1 kHz, they will detect the
BHMACHOs binaries before the Brazilian spherical detector. Note,
however that one spherical detector can determine the directions
of the sources (see Forward (1971) and also Magalh\~{a}es \etal
(1995, 1997) for details), while some bars and/or interferometers
are necessary to do the same.

In figure 1 we present the frequency of the GW as a function of
the coalescing time. We have used in this figure the well-known
paper by Peters and Mathews (1963) for a pair of 0.5~$\rm
M_{\odot}$ BHMACHOs.

Note that when the BHMACHO binaries are emitting GW at 100 Hz they
are about 10 seconds to coalescence . When this systems are
emitting GW at the frequency band of the bars they are a few
hundredths of second to coalescence. Finally, when the BHMACHO
binaries are emitting at the frequency band of the Brazilian
spherical detector they are $\sim$ 1 ms to coalescence.

It is worth stressing that, the BHMACHO binaries are ``chirping"
sources of GW when passing through the band of the
interferometers, the bars, and the spherical detectors such as the
Brazilian one.

In the following section, we consider the detectability of these
binary systems by the Brazilian spherical detector.

\section{Detectability of the BHMACHO binaries by the Brazilian spherical detector}
\label{sec-3}

\begin{figure}
\begin{center}
\epsfig{figure=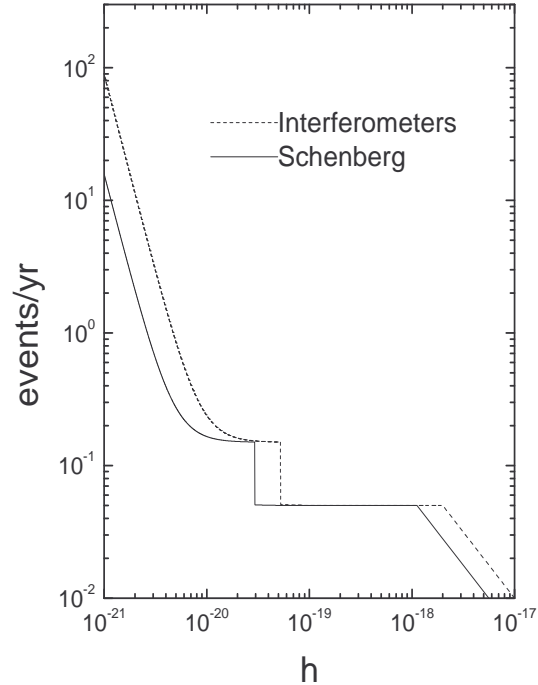,angle=360,height=11cm,width=8cm}
\caption{\label{fig2} Event rate, in events per year, as a
function of the sensitivity for interferometric detectors at 100
Hz, and for the Brazilian spherical detector at 3~kHz. We take
into account the following contribution: the Galaxy, M31, M33 and
the intergalactic BHMACHOs.}
\end{center}
\end{figure}

The detectability of the GWs emitted by BHMACHO binaries, by
either interferometers or resonant mass detectors, is easily
discussed in terms of the waves' ``characteristic amplitude''
$h_c$ (see, e.g., Nakamura \etal 1997, Thorne 1987):

\begin{equation}
h_c={\rm 4\times
10^{-21}\left(\frac{M_{chirp}}{M_\odot}\right)^{5/6}
\left(\frac{f_{GW}}{100Hz}\right)^{-1/6}\left(\frac{r}{20Mpc}\right)^{-1}}
\; ,
\end{equation}

\noindent where ${\rm M_{chirp} = (M_1 M_2)^{3/5} / (M_1 +
M_2)^{1/5}}$ is the ``chirp mass'' of the binary whose components
have individual masses ${\rm M_1}$ and ${\rm M_2}$; and ${\rm r}$
is the source-earth distance. The $h_c$ must be compared with a GW
detector's ``sensitivity to bursts''.

A BHMACHO binary of 0.5~$\rm M_{\odot}$ components, at 20 kpc to
the earth, is emitting GW at the Brazilian spherical detector band
with a characteristic amplitude of $h_{c}\sim 10^{-18}$. Since the
Brazilian spherical detector's sensitivity to burst is expected to
be $h_{s}\sim 10^{-20}$, the signal-to-noise ratio gives $SNR \sim
100$. Obviously the interferometers and the bars also detect such
systems, in particular the Galactic ones, at high values of
signal-to-noise ratios.

As seen, it would be very easy to detect BHMACHO binary systems,
but the main question here is how many of them are expected to be
detected, i.e., what is the event rate.

To assess the event rate related to the BHMACHO binary systems, we
follow Nakamura \etal (1997), who study the formation of such
systems. They consider that the BHMACHOs are part of the cold dark
matter, which is distributed throughout the universe.

Nakamura \etal obtain, in particular, the probability distribution
for the coalescence time for BHMACHO binaries. We refer the reader
to their paper for a detailed discussion of this issue.

Although Nakamura \etal derived the distribution function for the
coalescence, and we are dealing with the spiraling phase, in
particular the chirping phase, it is a good approximation to use
their results since we are at most, in the case of
interferometers, at 10 seconds of the coalescing phase.

For BHMACHOs in galaxies, such as the Galaxy, they find that the
event rate is $\sim$ 0.05 events/yr. Although the SNR for Galactic
BHMACHO binaries could be very high, it is not expected to detect
them easily.

It is easy to show, using equation (1), that the Brazilian
detector could be sensitive to BHMACHO binaries of the whole Local
Group ($r \sim 1.5$Mpc). But only M31 and M33 can give a
significant contribution to the event rate. The Galaxy, M31 and
M33 account for more then 90\% of the Local Group mass.
Considering that the contribution of M31 and M33 are similar to
that of the Galaxy, one would expect to detect with the Brazilian
detector 1 BHMACHO binary every 7 years.

Let us now consider the contribution of the BHMACHO binaries
distributed throughout the universe. Recall that the MACHO project
estimate that 20\% (with a 95\% confidence interval of 8\%-50\%)
of the Galactic halo could be in the form of BHMACHOs. We assume
that the BHMACHOs are $\sim$ 20\% of the dark matter too.

In figure 2 we show the event rate as a function of sensitivity
for interferometric detectors at 100 Hz, and for the Brazilian
spherical detector. We consider a signal-to-noise ratio equal to
unity. It is worth stressing that we are taking into account the
following contribution: the Galaxy, M31, M33 and the intergalactic
BHMACHOs, i.e., those distributed throughout the universe. Note
that the curve for the resonant bar detectors, which is not
plotted, would be located in between the curves for
interferometers and the Brazilian detector. The event rate for the
bars would be almost the same as the Brazilian detector.

Figure 2 has been constructed as follows. First of all instead of
using the event rate as a function of distance, we use equation 1
to obtain the event rate as a function of $h$ for a given
frequency and BHMACHO binaries' masses.

For distances below 20 $\textrm{kpc}$ the event rate increases
linearly from the center to the border of the Galaxy, where the
event rate is 0.05 events/yr. This corresponds to the linear
segment on the right side of the figure 2. From the border of the
Galaxy up to just before reaching M31 and M32 the intergalactic
BHMACHOs might be important, but since they follow the
distribution of the dark matter their contribution is negligible.
That is why there is a plateau in figure 2.

For a distance of 700-800 $\textrm{kpc}$, where M31 and M32 are
located, there is an additional contribution of 0.05 events/yr
from each of them. This corresponds to the step seen in figure 2.

For larger distances, say $>$ 1 $\textrm{Mpc}$, the background
BHMACHOs' contribution to the event rate begins to be relevant and
eventually the dominant contribution (see the left side of figure
2). Since the BHMACHO distribution is assumed to be homogeneous
throughout the intergalactic medium its contribution to the event
rate scales with the cubic power of the distance.

It is worth mentioning that to calculate the event rate related to
the intergalactic BHMACHO, it is necessary to take into account
the BHMACHO coalescing rate, which has been calculated using
equation (11) of the paper by Nakamura \etal (1997).

Finally, note also that if the Brazilian spherical detector's
sensitivity to bursts reaches $h_{s}\sim 10^{-20}-10^{-21}$, the
prospect of detecting BHMACHO binaries can be of several per year.

\section{Conclusions}
\label{conc}

We consider in the present paper the detectability of the putative
BHMACHO binaries by the Brazilian spherical detector.

We show that BHMACHO binaries of 0.5~$\rm M_{\odot}$ components
can be detected by the Brazilian detector at a very high SNR for
Galactic sources. However, the prospect for detecting such systems
in the Galaxy is at most 1 every 20 years.

Considering that such systems are ubiquitous among the galaxies,
and as a form of dark matter are distributed throughout the whole
universe, the prospect for their detection can be significantly
improved if the Brazilian spherical detector's sensitivity to
bursts reach $h_{s}\sim 10^{-20}-10^{-21}$.

Note that resonant bar detectors would see almost the same as the
Brazilian spherical detector, while the interferometers would see
a higher event rate, since they are more sensitivity to this kind
of signal than the other detectors.

Last, but not least, it worth mentioning, that one spherical
detector can determine the direction of the BHMACHO binaries,
while some bars and/or interferometers are necessary to do the
same.

\ack{ODM, ODA and BWP would like to thank the Brazilian agency
FAPESP for support (grants 02/07310-0 and 02/01528-4; 98/13468-9
and 03/04342-1; 02/07508-5, respectively). JCNA and ODA would like
to thank the Brazilian agency CNPq for partial financial support
(grants 304666/02-5, 300619/92-8, respectively). CSC would like to
thank the Brazilian agency CAPES for support. Finally, we would
like to thank the referee for useful comments and suggestions.}

\References

\item[]Alcock C \etal 1993 {it Nature} {\bf 365} 621

\item[]Alcock C \etal 2000 {\it Astrophys. J.} {\bf 542} 281

\item[]Aubourg E \etal 1993 {\it Nature} {\bf 365} 623

\item[]Burles  S, Nollett  K M and Turner M S 2001 {\it Astrophys.
J.} {\bf 552} L1

\item[]Forward R L 1971 {\it Gen. Rel. Grav.} {\bf 2} 149

\item[]Griest K 1991 {\it Astrophys. J.} {\bf 366} 412

\item[]Magalh\~{a}es N S, Johnson W W, Frajuca C and Aguiar O D
1995 {\it Mon. Not. R. Astron. Soc.} {\bf 274} 670

\item[]Magalh\~{a}es N S, Johnson W W, Frajuca C and Aguiar O D
1997 {\it Astrophys. J.} {\bf 475} 462

\item[]Nakamura T, Sasaki M, Tanaka T and Thorne K S 1997 {\it
Astrophys. J.} {\bf 487} L139

\item[]Paczy\'{n}ski B 1986a {\it Astrophys. J.} {\bf 301} 503

\item[]Paczy\'{n}ski B 1986b {\it Astrophys. J.} {\bf 304} 1

\item[]Peacock J 1999 {\it Cosmological Physics} (Cambridge:
Cambridge University Press)

\item[]Peters P C and Mathews J 1963 \PR {\bf 131} 435

\item[]Spergel D N \etal {\it Astrophys. J. Suppl.} {\bf 148} 175

\item[]Thorne K S 1987 {\it 300 years of Gravitation} (Cambridge:
Cambridge University Press) p~331

\item[]Udalski A, Szyma\'{n}ski M, Kalu\.{z}ny J, Kubiak M and
Mateo M 1992 {\it Acta Astron.} {\bf 42} 253

\item[]Yokoyama J 1977 {\it Astron. Astrophys.} {\bf 318} 673

\endrefs

\end{document}